\documentclass[sigconf]{acmart}
\AtBeginDocument{%
  }

\setcopyright{acmlicensed}
\copyrightyear{2018}
\acmYear{2018}
\acmDOI{XXXXXXX.XXXXXXX}
\acmConference[Conference acronym 'XX]{Make sure to enter the correct
  conference title from your rights confirmation email}{May 02,
  2025}{Brooklyn, NY}
\acmISBN{978-1-4503-XXXX-X/2018/06}

\usepackage{listings}
\usepackage{multirow}
\usepackage{graphicx}
\usepackage{soul}
\setcopyright{none}

\begin{document}

% \title{BLEAN: An LLM assisted Verilog code generation derived from CNF equations}
\title{Veritas: Deterministic Verilog Code Synthesis from LLM-Generated Conjunctive Normal Form}


\author{Prithwish Basu Roy}
\affiliation{%
  \institution{NYU Tandon School of Engineering}
  \city{Brooklyn, NY}
  \country{US}}
\email{pb2718@nyu.edu}

\author{Akashdeep Saha}
\affiliation{%
  \institution{NYU Abu Dhabi}
  \city{Abu Dhabi}
  \country{UAE}
}
\email{as19360@nyu.edu}

\author{Manaar Alam}
\affiliation{%
 \institution{NYU Abu Dhabu}
 \city{Abu Dhabi}
 \country{UAE}}
\email{alam.manaar@nyu.edu}

\author{Johann Knechtel}
\affiliation{%
  \institution{NYU Abu Dhabi}
  \city{Abu Dhabi}
  \country{UAE}}
\email{johann@myu.edu}

\author{Michail Maniatakos}
\affiliation{%
  \institution{NYU Abu Dhabi}
  \city{Abu Dhabi}
  \country{UAE}}
\email{michail.maniatakos@nyu.edu}

\author{Ozgur Sinanoglu}
\affiliation{%
  \institution{NYU Abu Dhabi}
  \city{Abu Dhabi}
  \country{UAE}}
\email{ozgursin@nyu.edu}

\author{Ramesh Karri}
\affiliation{%
  \institution{NYU Tandon School of Engineering}
  \city{Brooklyn, NY}
  \country{US}}
\email{rkarri@nyu.edu}

\renewcommand{\shortauthors}{Roy et al.}

\begin{abstract} 
Automated Verilog code synthesis poses significant challenges and typically demands expert oversight. Traditional high-level synthesis (HLS) methods often fail to scale for real-world designs. While large language models (LLMs) have enhanced scalability, they often introduce syntactical and logical errors requiring extensive post-generation verification. Here, we introduce a novel conjunctive normal form (CNF)-guided synthesis methodology. 
The idea is to have an LLM generate CNF clauses,
a format widely used for formal verification and synthesis validation in hardware design, but here it is used to
formally describe the desired circuit functionality.
These CNF specifications are then deterministically converted into Verilog, ensuring correctness by construction.
Our approach fine-tunes an open-source and lightweight LLM, namely the CPU-deployable LLama-3.2-3B-Instruct model (parameters < 4B), on a dataset of standard RTL components.
Experimental results demonstrate that our approach reliably produces functionally correct Verilog code on the first attempt, compared to other lightweight open-source SoTA works such as Verigen (2B parameters) and RTLCoder (4-bit quantized with around 7B parameters).
We will release our method and data in full post peer-review.
\end{abstract}






\maketitle

\section{Introduction} 
Designing hardware at the register-transfer level (RTL) in Verilog is complex and error-prone, traditionally requiring significant expertise~\cite{autochip}. To mitigate manual effort and increase abstraction, high-level synthesis (HLS) tools have been developed to translate high-level languages (e.g., C/C++) into hardware descriptions. Although HLS has been effective for simple circuits, it offers limited control over the resulting design and control paths. Further, verification continues to be a major bottleneck, with estimates indicating that over 50\% of development effort (in terms of cost) in ASIC/FPGA-based systems is spent on testing and debugging the RTL design~\cite{eval_LLM}.

\textbf{Emergence of LLMs for RTL Generation.} 
In recent years, there has been growing interest in LLMs for hardware design, with the objective of directly generating correct Verilog code from high-level specifications. This approach promises to significantly accelerate RTL development by enabling users to specify desired functionality in natural language while delegating low-level HDL coding to the LLM.

\textbf{Challenges for LLM-Driven RTL Generation.} 
Nevertheless, achieving correct-by-construction hardware with LLM assistance remains an open challenge.
In fact, early explorations of LLM-driven RTL design revealed further limitations. Code synthesized by an LLM often requires iterative refinement and verification. For example, AutoChip adopts a feedback-driven loop. It
combines a conversational LLM with a Verilog compiler and simulator to detect errors in the generated code. It then iteratively improves the HDL based on compilation errors or failing test cases~\cite{autochip}. This
workflow mirrors how a human engineer debugs code: generate an initial design, simulate it, fix the bugs, and repeat. In a recent study evaluating LLMs for hardware design and test, GPT-4 produced mostly functional Verilog but still needed human or tool interventions in roughly half of the test cases to meet the specifications~\cite{eval_LLM}.
\begin{figure}
    \centering
    \includegraphics[width=1.0\linewidth, trim=4cm 0 0 0, clip]{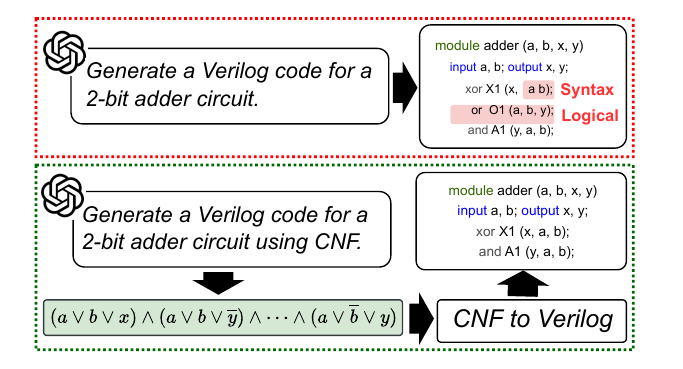}
    \caption{An overview of the proposed CNF-based Verilog code synthesis methodology.}
    % \vspace{-0.15cm}
    \label{fig:cover}
\end{figure}

\textbf{Limitations of the Generate‑then‑Verify Paradigm.} Notably, the LLM often struggled to create correct test benches, requiring guidance to verify its outputs~\cite{eval_LLM}. Such findings emphasize that current LLM-assisted workflows follow a ``generate-then-verify'' paradigm: the model's output must be checked by external means (simulation, test bench, or formal analysis) and then corrected in multiple passes if necessary. This dependence on external verification slows down the design cycle and demands additional tooling or expert oversight to ensure the RTL is correct.
 
\textbf{Recent Advances.}
Research has begun tailoring LLMs to hardware design by incorporating domain knowledge and integrating with electronic design automation (EDA) tools. ChatEDA, for instance, introduces AutoMage — a domain-specialized LLM that interfaces with CAD tools via APIs to autonomously identify and correct issues like syntax and timing errors within a tool-informed feedback loop~\cite{chateda}. In parallel, efforts like RTLCoder focus on improving model performance through training. RTLCoder fine-tunes a 7-billion-parameter model on over 27,000 synthetic Verilog problem-solution pairs to address the scarcity of high-quality HDL data~\cite{rtlcoder}. This results in significantly better RTL code generation, outperforming GPT-3.5 and approaching GPT-4-level accuracy on hardware design tasks. Despite these advances, current approaches still rely on external validation, such as simulation, test benches, or formal equivalence checking, to ensure correctness, reflecting a continued dependence on the ``generate-then-verify'' loop~\cite{autochip, eval_LLM}.

\textbf{Ours: CNF‑Guided Hardware Synthesis.} This work introduces a novel approach to LLM-assisted hardware design that explicitly breaks away from traditional generate-then-verify approach. We train the model to generate a formal hardware specification in the form of CNF Boolean clauses. CNF is a standard logical format consisting of a conjunction of one or more disjunctions (clauses) of Boolean literals. It can precisely encode an RTL design’s truth table or Boolean function in a manner amenable to algorithmic manipulation. Instead of directly writing Verilog code, which may contain subtle errors (syntactical or logical), the model generates a CNF specification for a new design prompt. We then synthesize a Verilog implementation from that CNF using a deterministic algorithm (refer  Figure~\ref{fig:cover}). Since the CNF generated by LLM precisely encodes the functional requirements (and can be derived from a known correct design or specification), the translated Verilog is guaranteed to satisfy those requirements by construction. In essence, verification step is integrated into the generation process. If the CNF is satisfiable and aligns with intended functionality, the resulting hardware will logically meet the specification.

\textbf{Our Approach and Contributions.}
Our approach involves collecting baseline designs (adders, decoders, subtractors, multiplexers) and converting each into CNF formulae that characterize its correct input-output behaviour. These serve as training data for LLM, which is fine-tuned to output CNF clauses, capturing the desired logic of circuit given high-level descriptions. Consequently, the LLM operates at the level of formal logic rather than syntactic code. This CNF-guided workflow eliminates the need for separate test bench validation or formal proof after code generation, as the LLM’s output is a verifiable specification itself. The resulting RTL generation pipeline ensures correctness a priori, clearly contrasting with previous approaches that discover correctness a posteriori through expert simulation or debugging. In summary, Veritas
makes the following contributions.
\begin{itemize}
\item  We propose Veritas, a novel method for CNF-guided Verilog synthesis that leverages LLMs to generate hardware in a correct-by-construction manner, by outputting formal CNF representations that are later converted into Verilog. To the best of our knowledge, this is the first approach that tightly couples an LLM's output with a formal hardware synthesis step, merging specification and implementation.
\item We develop a complete training and translation pipeline starting from a generic set of Verilog designs and their CNF equivalents, we fine-tune a transformer-based model to accurately produce CNF clauses for target RTL functionalities and implement an automated system that converts these clauses into synthesizable Verilog code.
\item We demonstrate through case studies and \emph{pass@k} metrics comparison with other open-source lightweight models like Verigen-2B and RTLCoder, that our CNF-trained LLM can generate non-trivial RTL components (e.g., an ALU) that operate correctly on the first attempt without iterative debugging. Compared to conventional LLM-assisted design, our approach eliminates the need for external verification (no post-generation test benches or formal checkers). 
\textit{We will open-source all the code and data post peer-review.}
 
\end{itemize}
% 
%The rest of the paper is organized as follows...
% 
\section{Background and Motivation}
The current state-of-the-art generative models like LLama 3, GPT, DeepSeek, etc., which are trained on enormous data corpora, can comprehend prompts in natural language and produce human-like responses. Although they are commendable for generic tasks, the limitation of the data corpus on which they have been trained restrains them from answering very detailed domain-specific queries.
\subsection{Training Generative Models}
The domain experts resort to the following techniques to cater to the finer needs of domain-specific response generation.
\subsubsection{Prompt Engineering} In this technique, the user tweaks the prompt appropriately to obtain optimal responses from the model. This technique is quite cost-effective and does not require additional training
in the vanilla LLM. Prompt engineering often involves breaking down a complex problem into small, easily solvable subproblems. Works like ChipChat~\cite{chipchat}, use prompt engineering and a human hardware design
expert in a loop to generate an 8-bit accumulator-based microprocessor. The works \cite{eval_LLM} and \cite{autochip} remove human intervention by introducing a feedback framework that evaluates the responses and provides feedback to the model to improve the quality of the response.

\subsubsection{In-Context Learning} This special case of prompt engineering involves zero-shot, one-shot, and few-shot learning. Models like GPT4, thanks to their vast training corpus, can be directly asked domain-specific questions. Thanks to their large context size, they can often provide acceptable responses to the never-before-seen prompts. This scenario is called zero-shot learning. In one-shot and few-shot learning, the user generally provides one or multiple examples of an expected response in the prompt and expects the model to learn from those sample instances. This method might not be suitable for tasks that require complex understanding of the relation of diverse components.

\subsubsection{Fine-Tuning} The pre-trained LLM model must be trained on a curated domain-specific dataset to address domain-specific tasks. Since training a complete LLM model involves tweaking billions of parameters,
	it is time-consuming and highly compute-intensive. Experts proposed solutions such as Low-Rank adaptations of LLMs (LoRa~\cite{LoRA}) to tackle this scenario. LoRa preserves the weights of the pre-trained model
	and appends a significantly small trainable matrix to each layer of the transformer architecture. The uniqueness of this added matrix lies in its decomposability into two low-rank matrices, whose weights get
	updated on fine-tuning. The low dimensionality of these matrices allows LoRa to be highly efficient in computing, i.e., without introducing significant latency. ChatEDA~\cite{chateda} fine-tunes LLama2 using LoRa to assist the end-to-end RTL to GDSII flow. Verigen~\cite{thakur2024verigen} performs instruction-tuning on Codegen 16B using a curated dataset comprising RTL descriptions as prompts and the corresponding RTL Verilog code as completions. 

\subsubsection{Control Parameters for Generative Tasks} An LLM's uniqueness lies in creating a dictionary of possible following tokens and their corresponding probability of occurring next, given the current context (set of preceding $n$ words/tokens). LLMs allow the user to vary the randomness of the prediction of the next token by tweaking two parameters, \textit{temperature} and \textit{top\_p}. The \textit{temperature} parameter adjusts the randomness of the output. The low temperature value provides a deterministic outcome, while the high temperature allows more randomness in token generation. The \textit{top\_p} parameter forces the LLMs to consider the \textit{top\_p} tokens from the possible following tokens. These parameters aid in generating quality output post fine-tuning phase.

\subsection{Conjunctive Normal Form (CNF)}\label{subsec:cnf}
CNF is a standardized representation of Boolean formulas as an AND of clauses, each of which is an OR of literals (variables or their negations). This structured form has long been associated with ensuring the correctness of circuit representations~\cite{seltner2014extracting}.

One widely adopted method for generating a CNF from a circuit is the Tseytin transformation~\cite{tseitin}, which converts any combinational logic circuit into an equisatisfiable CNF formula. By assigning auxiliary (Tseytin) variables to each gate and encoding each gate's functionality as a set of CNF clauses, this transformation ensures that the overall formula grows linearly with the circuit's size while preserving each component's operational integrity.
This CNF representation is particularly effective for SAT solver-based verification. SAT solvers excel at resolving Boolean satisfiability problems, making them ideal for formal verification tasks in EDA~\cite{matsunaga1996efficient}.

A prominent application is combinational equivalence checking (CEC) using so-called miter circuits~\cite{brayton1990multilevel}. In a miter circuit, the outputs of two circuits are compared, often through XOR gates, to capture differences, which are then aggregated into a single signal~\cite{miter}. Transforming this miter circuit into CNF using the Tseytin transformation allows SAT solvers to determine equivalence efficiently: an unsatisfiable result confirms functional equivalence, while a satisfiable result yields a counterexample~\cite{mishchenko2006improvements}.

Moreover, the association of CNF with circuit correctness extends to hardware synthesis. A circuit accurately represented in CNF can be systematically translated into a Verilog description, ensuring the resulting hardware description is correct by construction. This integration of formal methods with practical design automation underpins reliable and efficient EDA workflows where each gate and the circuit as a whole meets its intended specifications.
\subsection{Why CNF in General and Propositional Logic Formula (PLF) in Particular?}
\begin{figure}
    \centering
    \includegraphics[width=1.05\columnwidth, trim=0cm 0cm 0cm 0cm, clip]{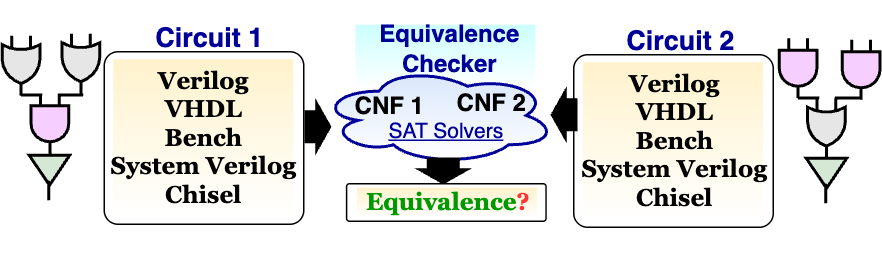}
    \caption{The SAT solvers utilize CNF representation of circuits to check equivalence.}
    % \vspace{-0.15cm}
    \label{fig:why_cnf}
\end{figure}

Directly using circuit netlists for fine-tuning LLMs may seem straightforward. However, it presents several challenges. Netlists derived from RTL can exhibit significant syntactic variability due to different design styles and language constructs. In contrast, CNF offers a standardized and uniform representation of circuit behaviour, making it a more suitable intermediate format. Further, formal equivalence checking tools (ABC~\cite{abc} and Formality~\cite{formality}) internally rely on CNF representations through miter constructions for SAT-based verification (refer Figure~\ref{fig:why_cnf}). Leveraging CNFs for LLM fine-tuning aligns with these formal methods and mitigates syntactic and semantic inconsistencies commonly observed in LLM-generated RTL. A post-processing agent can then translate LLM-generated CNFs into functionally correct RTL, thereby improving reliability and correctness of the overall synthesis pipeline.

We utilize PLF during fine‑tuning to reduce token overheads, but our ultimate goal remains to produce CNF representations. PLF combines logical operators like AND, OR, NOT, implication $(\rightarrow)$, and bi-implication $(\leftrightarrow)$. 
Refer to Eq.~\ref{eq:PLF} for PLF and Eq.~\ref{eq:CNF} for CNF representation using Tseytin transformation of $X=XOR(A, B)$. 

\begin{gather}  \label{eq:PLF}
    X \leftrightarrow \overline{(A \leftrightarrow B)}
\end{gather}
\begin{gather}  \label{eq:CNF}
    (\overline{A}\vee \overline{B}\vee \overline{X})\wedge(A\vee B \vee \overline{X})\wedge(A\vee \overline{B} \vee X)\wedge(\overline{A} \vee B\vee X)
\end{gather}

See Table~\ref{tab:case-study1} for CNF representation of other gates.
Any PLF can be transformed into an equisatisfiable CNF using a standard, linear-time Tseytin construction~\cite{huth2004logic}. This conversion can increase formula size and, thus, token count. Adopting PLFs for the initial tuning stage lets us optimize model costs without compromising our ability to generate the full CNF encodings required by downstream SAT‑based equivalence checks~\cite{abc, formality}.

\subsection{The pass@k Metric} The metric \textit{pass@k} is commonly used in measuring the success rate of generative tasks tackled by an LLM. If an LLM is queried $n$ times, then \textit{pass@k} is given by the probability of at least one LLM response being correct out of $k$ samples, when a total $c$ out of the $n$ samples are correct. Formally, it is defined by as follows.
\begin{equation}
    pass@k = 1 -  \frac{^{n-c}C_{k}}{^nC_k}
\end{equation}
where $^{m}C_k$ signifies the number of ways a sample set of size $k$ can be selected from a set of $m$ elements.

The \textit{pass@k} metric is widely accepted for for evaluating code generation. It has been used in VerilogEval~\cite{VerilogEval} and RTLLM~\cite{RTLLM} who check for the probabilty of generation of one piece of functionally correct code for $k$ given prompts out of $n$ prompts. In our work we use \textit{pass@k} to check the success rate of CNF generation as well as CNF-assisted Verilog code generation. 

\section{Related Works}
This section refers to some essential prior work, beginning with methods that leverage prompt engineering for Verilog generation and then discussing approaches based on fine‑tuning.
\subsection{Verilog Coding with Prompt Engineering}
In ChipGPT \cite{chipGPT}, the authors try to refine LLM's Verilog code generation by appropriately pre-processing prompts. By providing step-by-step feedback about the generated code, they improve code generation
incrementally. Post-generation, they do a power-performance-area analysis of different variants of the same code and provide the optimal solution. In~\cite{chipchat}, the authors used prompt engineering to generate RTL
code using GPT-4. With a significant involvement of human feedback, the authors could design a tape-out-able microprocessor. In~\cite{eval_LLM}, the authors evaluated the design and test generation capability of various open-source and closed-source models. They categorized the performance of models based on the intensity of human effort required to get Icarus Verilog (iVerilog)-compilable and functionally correct designs. The authors in AutoChip~\cite{autochip} provided a framework in which the LLM receives compilation and simulation feedback from iVerilog~\cite{iVerilog}. This feedback is appended to the succeeding prompts in case of an incorrect generation. In `full' feedback mode, the feedback from all the previous iterations is appended to the current prompt, while for the `succinct' setup, only the last two feedbacks are only considered. With the inclusion of such feedback, AutoChip reported a 21.19\% increase in the code generation accuracy compared to zero-shot prompts provided to GPT-4. For open-source models like Claude 2, the improvement was limited to~15\%. AutoChip fails if the generation and feedback loop crosses 10 iterations and does not provide a correct output. 
\subsection{Verilog Coding with Instruction Tuning}
In DAVE \cite{DAVE}, the authors, for the first time, fine-tuned a GPT2 model to generate Verilog code. In Verigen \cite{thakur2024verigen}, the authors created a dataset using the Verilog codes publicly available in texts on websites and textbooks, and fine-tuned Codegen-2B and Codegen-16B. Although Verigen could beat the benchmarks of ChatGPT3.5, its overall code generation capability remained poor, often generating incomplete and buggy code whenever the design requirement is a bit high or the prompt is complex. 
Using non-pre-processed raw data without quality evaluation for training was mainly responsible for the poor code generation. In RTLCoder, the authors fine-tuned Mistral-7B-v0.1 and DeepSeek-Coder-6.7b-v1 with their
custom dataset and showed that their fine-tuned model significantly improved (5-10\%) the  \emph{pass@k} rates from the base models. Apart from syntax checks of the training data, RTLCoder also conducts a functionality check, ensuring correct and high-quality code generation. They outperform GPT3.5 on the VerilogEval framework and are comparable to GPT4 in performance. 
 
\begin{figure}
    \centering
    \includegraphics[width=1.0\columnwidth, trim=1.4cm 0.0cm 2.6cm 0cm, clip]{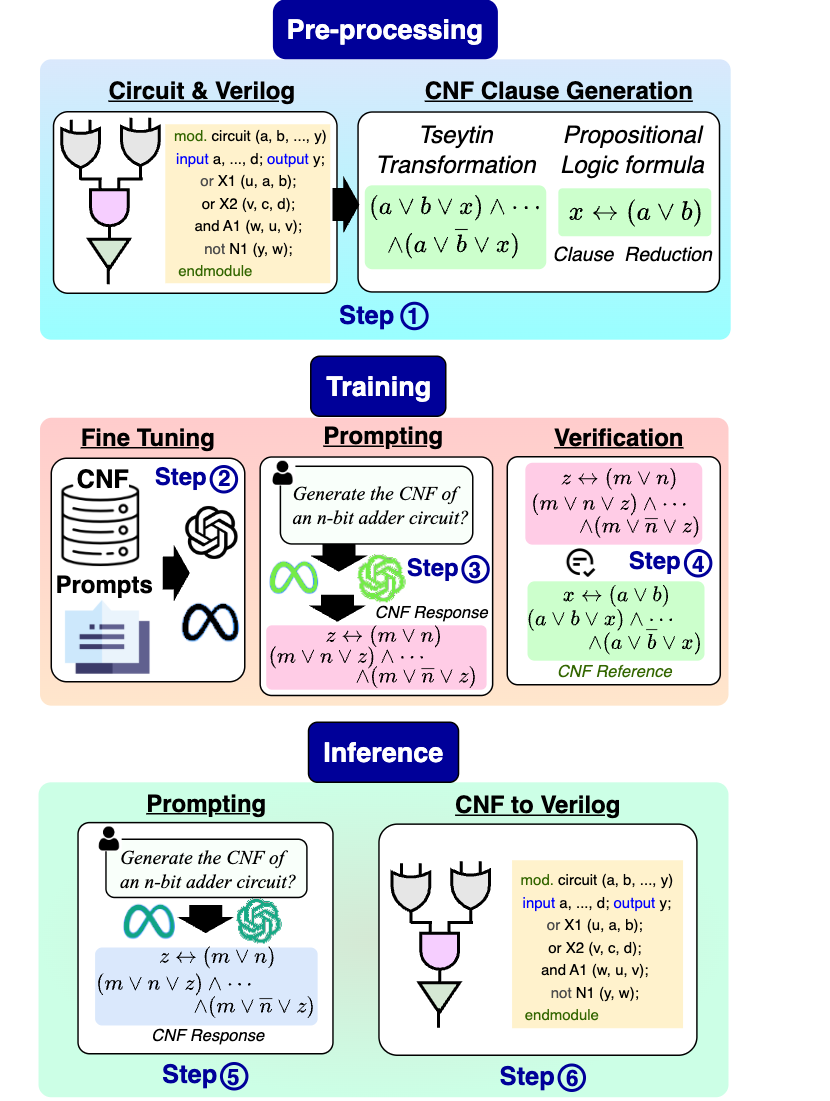}
    \caption{An overview of steps involved in the proposed Veritas framework.}
    % \vspace{-0.15cm}
    \label{fig:method}
\end{figure}
\section{Methodology}
In this section, we propose our method, Veritas, a fine-tuned LLM model capable of generating CNF equivalence clauses for an unseen generic design. 
Our work differs from prior art in one critical aspect -  we do not train the LLM on some curated set of Verilog codes. Instead, we train our
model with CNF equations of commonly used circuits. We leverage the CNF-generating capabilities of the LLM to generate correct Verilog code deterministically. 
 
\subsection{Veritas Components}
Next, we describe the components of Veritas in detail. Note that the data pipeline is described separately further below.
 
{\bf The CNF clause generator.} Given a gate-level netlist, the generator can generate a set of CNF clauses representing the whole circuit.
%We use the Synopsys Design Compiler to generate the gate-level netlist of a given RTL Verilog code.\jk{confusing/misleading here; not essential anyway}
A gate-level netlist is a gate-wise representation of a given digital circuit. Since each of the basic gates has a corresponding CNF representation with Tseytin transformation~\cite{tseitin}, a gate-level netlist can be represented as a conjunction of the CNF equations of the individual gates (refer \textbf{Step \textcircled{1}} in Figure ~\ref{fig:method}).
 
{\bf The equivalence clause reduction.} 
The Tseytin transformation-based CNF encoding of basic logic gates typically results in 2 to 4 clauses per gate, depending on the gate type. While this approach facilitates a straightforward translation of circuits into CNF, it does not scale efficiently for large, real-world designs. To address this scalability challenge and reduce the clause count, we adopt a PLF for each gate. Note thate the conversion from PLF to CNF is standard and straightforward~\cite{huth2004logic}.

\begin{figure}
    \centering
    \includegraphics[width=1.0\columnwidth]{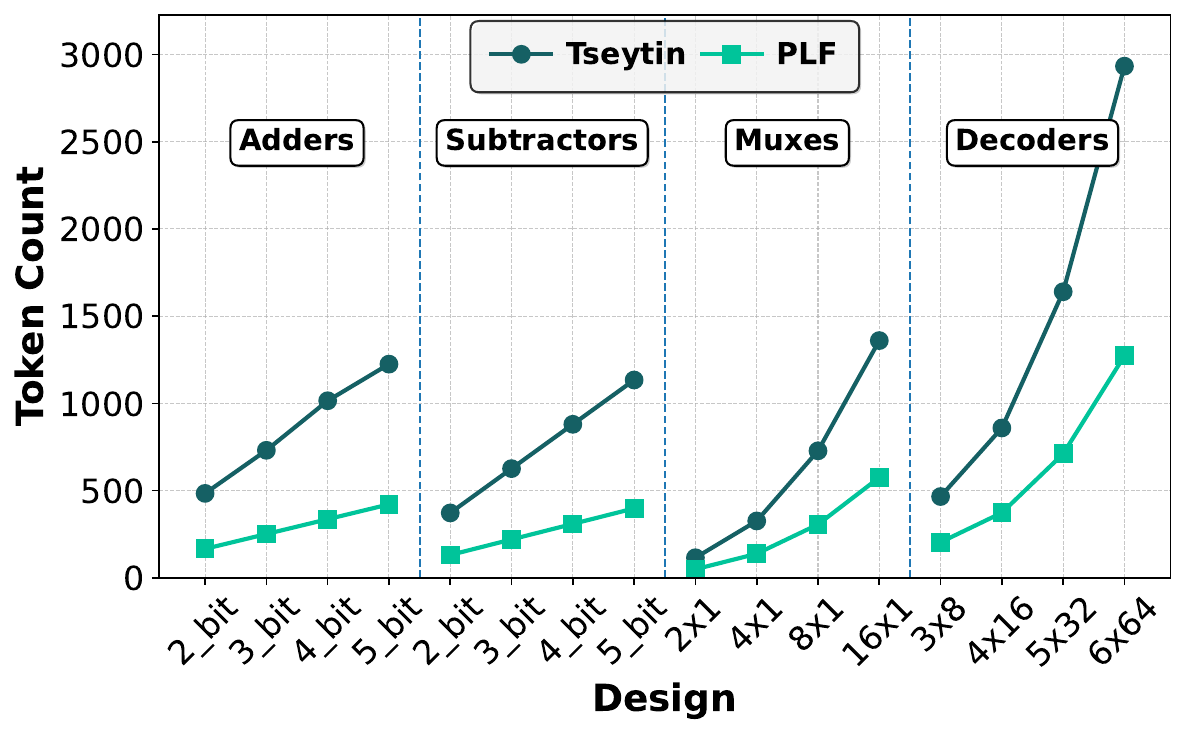}
    \caption{PLF requires considerably fewer tokens compared to Tseytin transformation-based CNF representation. The gap between the number of tokens required increases significantly as the size of the designs increase from a simple adder to a complex decoder.}
    % \vspace{-0.15cm}
    \label{fig:plfvstseytin}
\end{figure}

This intermediate representation substantially reduces clause size, as shown in Table ~\ref{tab:clause_reduc}, thereby significantly lowering the number of tokens required for LLM fine-tuning. We further show the effectiveness of using PLF expressions in place of Tseytin transformation in Figure~\ref{fig:plfvstseytin}. Tseytin transformation of a $6\times64$ decoder takes $2.30\times$ more tokens than its equivalent PLF representation, while for a 5-bit adder $2.98\times$ more tokens are required.
Note how \textbf{Step \textcircled{1}} shows that our framework has the capability of generating both Tseytin Transformations and Proposition Logic Formulas. We utilize primarily the PLF in our work; however, CNFs are utilized for equivalence checking utilizing SAT solvers~\cite{abc,formality}.

\begin{table}[b]
\caption{Reduction of the size of the clauses for basic gates compared to Tseytin transformation (TT) vs PLF.}
\scalebox{0.72}{
\begin{tabular}{|c|c|c|c|c|c|c|c|c|}
\hline
\textbf{Gates} & \textbf{AND} & \textbf{NAND} & \textbf{OR} & \textbf{NOR} & \textbf{BUF} & \textbf{NOT} & \textbf{XOR} & \textbf{XNOR} \\ \hline
\textbf{TT} & 3 & 3 & 3 & 3 & 2 & 2 & 4 & 4 \\ \hline
\textbf{PLF} & 1 & 1 & 1 & 1 & 1 & 1 & 1 & 1 \\ \hline
\textbf{Reduc. (\%)} & \textbf{66.67} & \textbf{66.67} & \textbf{66.67} & \textbf{66.67} & \textbf{50} & \textbf{50} & \textbf{75} & \textbf{75} \\ \hline
\end{tabular} }
\label{tab:clause_reduc}
\end{table}
 
{\bf A fine-tunable base LLM model.} We choose LLama3.2-3B-Instruct as our open-source base model. LLama3.2 has been trained on a vast corpus of open-source data and has probably been exposed to CNF equations in its pre-training data.
%To ensure this is not the case,\jk{misleading}
We queried the base model to validate it to give us the CNF equations for the basic gates namely AND, OR, NAND, NOR, XNOR, XOR, NOT, and BUF.

Although the model realized what a CNF equation is, it mostly returned Boolean expressions of the logic gates. Following this observation, we decided to fine-tune the base model with the CNF equations for all the basic gates before further fine-tuning it with the CNF representations of various generic designs.
 
{\bf The CNF clause equivalence checker.} The fine-tuned LLama model can generate CNF clauses for unseen designs. To evaluate the accuracy of the model's CNF equation generation capability, we compare the generated CNFs with the golden CNFs. 

While comparing the two CNFs, there are a couple of challenges. First, the equivalence checker considers the possibility that the position of clauses representing the gates might change. Second, the name of the intermediate wires can also vary. 
We have designed our equivalence checker to overcome these challenges to compare two CNFs. If the CNFs are not comparable, it informs the user about the possible mismatch regarding the number of clauses or differences in a particular gate logic. We use the CNF clause equivalence checker in the verification phase of the training (refer \textbf{Step \textcircled{4}} in the training phase of Figure \ref{fig:method}).
 
{\bf CNF to Verilog convertor.} The generated CNFs in the inference phase (\textbf{Step \textcircled{5}}) are first converted to an equivalent bench file using the `ABC' tool~\cite{abc}. The bench files are then converted to any equivalent RTL file using simple deterministic scripts (\textbf{Step \textcircled{6}}).

\subsection{Veritas Data Pipeline} 
The Veritas data flow/pipeline consists of three stages:
the pre-processing, the training, and the inference stages. We explain each of these stages in detail below.

{\bf Pre-processing stage.} The pre-processing phase involves the preparation of training data. At first, gate-level synthesis of logical designs comprising eight basic gates, 2 to 4-bit adders, 2 to 4-bit subtractors,
	$2\times1$ to $16\times1$ multiplexers, and $2\times4$ decoders to  $5\times32$ decoders is performed following different design constraints. By enforcing design constraints, we achieve different gate-level variants of the same design. Different variants lead to different but equivalent Tseytin or PLF representations of the same design.

We prepare our dataset by combining descriptions of logical designs mentioned above and their respective Tseytin transformations and PLF equations. The training data preparation is depicted in \textbf{Step \textcircled{1}} of Figure \ref{fig:method}.

{\bf Training.} The training dataset from \textbf{Step \textcircled{1}} is partitioned into 80\%, 10\%, and 10\% train, validation, and test split. We use LoRa to fine-tune the model and fine-tune it for 20 epochs and a learning rate of $1e-03$, with a LoRa rank of 16 (\textbf{Step \textcircled{2}}). After each epoch, we verify the generated CNFs with the reference CNFs from the validation set (\textbf{Step \textcircled{3}}). We stop the fine-tuning when the training and validation loss stabilizes. 
We do an additional verification round with unseen test data (\textbf{Step \textcircled{4}}), where the CNF equivalence checker compares the generated CNF equivalence clauses with the golden CNF equivalence clauses. Based on the evaluation, we decide whether we want to fine-tune further.

{\bf Inference.} In \textbf{Step \textcircled{5}}, we test the fine-tuned model from the training stage with previously unseen prompts for designs like a 5-bit adder, 5-bit subtractor, 6$\times$64 decoder, etc.
\textbf{Step \textcircled{6}} converts the CNF clauses to .bench format and subsequently to a Verilog RTL file using the `ABC' tool.

\section{Empirical Validation}

\subsection{Experimental Setup}
We fine-tuned our LLama-3.2-3B-Instruct model on an Nvidia-A100-80 GB GPU.
We arrange for multiple case studies as follows.
For Case Study 1, we prepared the dataset comprising eight basic gates with 131 training data points, 19 validation data points, and 18 test data points. For Case Study 2, we prepared the dataset as a combination of the description 2 to 4-bit adders, 2 to 4-bit subtractors, $2\times1$ to $16\times1$ multiplexers, and $2\times4$ decoders to  $5\times32$ decoders and their corresponding CNFs as completions.
We used a training dataset comprising of 1049 points, a validation dataset of 131 points, and a test dataset comprising unseen prompts of 132 data points. 
We use the fine-tuned model from Case Study 2 to generate an ALU from scratch and for the SoTA comparison as in Case Study 3. 

\begin{figure}[t]
    \centering
    \includegraphics[width=1\linewidth]{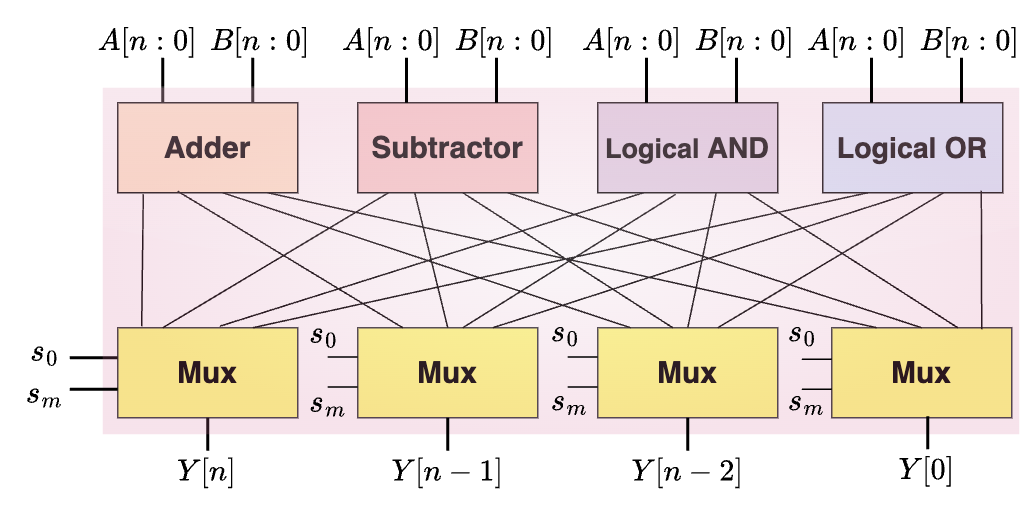}
    \caption{Block diagram of a fully functional ALU designed from scratch using Veritas.}
    \label{fig:toy_alu}
\end{figure}

\subsection{Case Study 1} 

\textbf{Scope.}
In Section~\ref{subsec:cnf}, we established that CNFs can be easily derived from gate-level representation of an RTL design (say represented using Verilog) and vice-versa. In this case study, we analyze \emph{the basic CNF-generating capabilities} of current commonly used LLMs.
\begin{lstlisting}[caption={Incorrect logic of 4-bit Combinational Subtractor, as generated by RTLCoder.}, label={lst:4bit_sub_RTLCoder}]
§Prompt:§
Please act as a professional Verilog designer.

Implement a 4-bit subtractor for calculating the difference and borrow of two 4-bit numbers
§Module name:§
    4_bit_subtractor
§Input ports:§
        a_3_ : 3rd bit of input a
        a_2_ : 2nd bit of input a
        a_1_ : 1st bit of input a
        a_0_ : 0th bit of input a
        b_3_ : 3rd bit of input b
        b_2_ : 2nd bit of input b
        b_1_ : 1st bit of input b
        b_0_ : 0th bit of input b
§Output ports:§
        diff_3_ : 3rd bit of output diff
        diff_2_ : 2nd bit of output diff
        diff_1_ : 1st bit of output diff
        diff_0_ : 0th bit of output diff
        borrow :  borrow bit
§Implementation:§
The implementation is a combinational subtractor logic with a 4 bit input a, and a 4 bit input b. The output is 4 bit. borrow is 1 bit
Give me the complete code.

module 4_bit_subtractor(
    input      a_3_,
    input      a_2_,
    input      a_1_,
    input      a_0_,
    input      b_3_,
    input      b_2_,
    input      b_1_,
    input      b_0_,
    output     diff_3_,
    output     diff_2_,
    output     diff_1_,
    output     diff_0_,
    output     borrow
);

§Completion:§
assign {borrow, diff_0_, diff_1_, diff_2_, diff_3_} = {~a_0_^b_0_, ~(a_0_^b_0_) & (a_1_^b_1_), ~(a_1_^b_1_) & (a_2_^b_2_), ~(a_2_^b_2_) & (a_3_^b_3_), a_3_ ^ b_3_};
endmodule

§Evaluation§:
§A = 0000, B = 0000 => Diff = 0000, Borrow = 1§
A = 1000, B = 0100 => Diff = 1010, Borrow = 1
A = 0111, B = 0001 => Diff = 0001, Borrow = 1
A = 0011, B = 0100 => Diff = 0000, Borrow = 0
\end{lstlisting}
\textbf{Setting.}
We selected four open-source models, namely Mistral-7B~\cite{mistral7b}, DeepSeek-R1-7B~\cite{deepseek-coder}, LLama 3.2-3B-Instruct~\cite{llama3p2_3b_instruct}, and LLama3.1 8B~\cite{llama3p1_8b}. We also chose the latest commercial OpenAI model, GPT-4o-mini~\cite{GPT4o-mini}, and the two other legacy models, GPT-4~\cite{GPT4} and GPT-3.5-Turbo~\cite{GPT3.5-turbo}. For this basic test, we prepared a set of 20 prompts, out of which the first five prompts asked the models to generate CNF for the NAND gate, the next five prompts asked for the CNF generation of the NOR gate, 
the following five prompts asked to generate CNF equations for the XOR gate, and finally, the last 5 prompted the equations for buffer.
We calculated the \emph{pass@k} metric for $k = 1, k = 5,$ and $k = 10$, while $n$ was fixed at twenty. 

\begin{table*}[]
\caption{Comparison of generated CNFs for basic gates in various models. The highlighted row shows correct CNFs generated compared to the original Tseytin transformation (first row) for the gates.}
\label{tab:case-study1}
\renewcommand{\arraystretch}{1.5}
 \resizebox{2.1\columnwidth}{!}{%
\begin{tabular}{|c|c|c|c|c|}
\hline
\textbf{Models} & \textbf{NAND} & \textbf{NOR} & \textbf{XOR} & \textbf{BUF} \\ \hline
\textbf{Tseytin~\cite{tseitin}}  & $(\overline{A}\vee \overline{B}\vee Y)\wedge(A\vee \overline{Y})\wedge(B\vee \overline{Y})$ & $(A\vee B\vee Y)\wedge(\overline{A}\vee \overline{Y})\wedge(\overline{B}\vee \overline{Y})$ & \begin{tabular}[c]{@{}l@{}}$(\overline{A}\vee \overline{B}\vee \overline{Y})\wedge(A\vee B \vee \overline{Y})$\\ $\wedge(A\vee \overline{B} \vee Y)\wedge(\overline{A} \vee B\vee Y)$\end{tabular} & $(A \vee \overline{Y})\wedge(\overline{A} \vee Y)$ \\ \hline
\textbf{Mistral-7B~\cite{mistral7b}} & $(A \vee \overline{B}\vee Y)\wedge(A\vee B \vee \overline{Y})$ & $(\overline{A} \vee \overline{B}\vee \overline{Y})\wedge(\overline{A}\vee B \vee Y)$ & \begin{tabular}[c]{@{}l@{}}$(A\vee \overline{B}\vee \overline{Y})\wedge(\overline{A}\vee B \vee \overline{Y})$\\ $\wedge(A\vee B \vee \overline{Y})\wedge(\overline{A} \vee \overline{B} \vee Y)$\end{tabular} & $(\overline{A} \vee \overline{Y}) \wedge (B \vee Y)$ \\ \hline
\textbf{DeepSeek-R1-7B~\cite{deepseek-coder}}& $( \overline{A} \vee \overline{B})$ & $( \overline{A} \vee \overline{B})$ & $( A \vee B) \wedge ( \overline{A} \vee \overline{B})$ & $(A \vee \overline{Y})$ \\ \hline
\textbf{LLama3.2-3B-Inst.~\cite{llama3p2_3b_instruct}} & $( \overline{A} \wedge \vee \overline{B}) \wedge (\overline{A})$ & $(A \wedge \vee B) \vee (\overline{A} \wedge \vee  \overline{B} )$ & $( A \wedge \vee \overline{B}  \wedge (Y \wedge \vee \overline{Y})$ & $( \overline{A} \vee Y) \wedge (\overline{Y} \vee B)$ \\ \hline
\textbf{GPT-3.5 Turbo~\cite{GPT3.5-turbo}} & $( A \vee B \vee \overline{Y}) \wedge (\overline{A} \vee \overline{B} \vee Y)$ & $(\overline{A} \vee \overline{B} ) \wedge Y$ & $(A \vee Y) \wedge (\overline{A} \vee \overline{B} )$ & $(A \vee \overline{Y} ) \wedge (A \vee Y)$ \\ \hline
\textbf{GPT4~\cite{GPT4}} & $(Y \vee A) \wedge ( Y \vee B)$ & $(A \vee Y)\wedge(B \vee Y)\wedge( A \vee \overline{B} \vee \overline{Y} )$ & \begin{tabular}[c]{@{}l@{}}$(\overline{A}\vee \overline{B}\vee \overline{Y})\wedge(A\vee B \vee \overline{Y})$\\ $\wedge(A\vee \overline{B} \vee Y)\wedge(\overline{A} \vee B\vee Y)$\end{tabular} &  \textcolor{green}{$\mathbf{(A \vee \overline{Y}) \wedge(\overline{A} \vee Y)}$} \\ \hline
\textbf{GPT-4o-mini~\cite{GPT4o-mini}} & $( A \vee B \vee \overline{Y}) \wedge (\overline{A} \vee \overline{B} \vee \overline{Y})$ & \begin{tabular}[c]{@{}l@{}}$(\overline{A} \vee \overline{B}) \wedge (Y \vee A) \wedge (Y\vee B)$ \\ $\wedge (\overline{Y}\vee \overline{A})  \wedge \overline{A}\vee \overline{B}$\end{tabular} & $(A \vee B) \wedge (\overline{A}\vee \overline{B})$ & $(A \vee Y) \wedge (\overline{A} \vee \overline{Y} )$ \\ \hline
\textbf{Veritas [Ours]} &  \textcolor{green}{$\mathbf{(\overline{A}\vee \overline{B}\vee Y)\wedge(A\vee \overline{Y})\wedge(B\vee \overline{Y})}$} & \textcolor{green}{$\mathbf{(A\vee B\vee Y)\wedge(\overline{A}\vee \overline{Y})\wedge(\overline{B}\vee \overline{Y})}$} &  \begin{tabular}[c]{@{}l@{}}\textcolor{green}{$\mathbf{(\overline{A}\vee \overline{B}\vee \overline{Y})\wedge(A\vee B \vee \overline{Y})}$}\\ \textcolor{green}{$\mathbf{\wedge(A\vee \overline{B} \vee Y)\wedge(\overline{A} \vee B\vee Y)}$}\end{tabular} & \textcolor{green}{$\mathbf{(A \vee \overline{Y})\wedge(\overline{A} \vee Y)}$}\\ \hline
\end{tabular} 
} 
\end{table*}

\textbf{Results.}
As evident from Table~\ref{tab:case-study1}, most of the generated CNFs are incorrect, while LLama 3.2-3B-Instruct (our base model) produces syntactically incorrect CNFs for different gates. Our fine-tuned model (Veritas) generates correct CNF for all the basic gates. Similar behaviour is also observed with models producing the PLFs.
Table~\ref{tab:baseline} gives a quantitative analysis of the CNF generation capabilities of various open~\cite{ollama} and closed-source models. Among the models, only GPT4 generated the correct output every time for buffer(BUF) and only once for XOR, and thus has a high pass@k.

% \usepackage{multirow}
% \usepackage{graphicx}

% % Please add the following required packages to your document preamble:
% \usepackage{graphicx}
% \usepackage{graphicx}
\begin{table}[]
\caption{Performance of various pre-trained LLM models compared to our fine-tuned LLama3.2 in generating the CNF of basic gates.}
\label{tab:baseline}
\resizebox{\columnwidth}{!}{%
\begin{tabular}{|c|c|c|c|}
\hline
\textbf{Models} &
  \multicolumn{1}{c|}{\textbf{pass@1}} &
  \multicolumn{1}{c|}{\textbf{pass@5}} &
  \multicolumn{1}{c|}{\textbf{pass@10}} \\ \hline
\textbf{Mistral-7B~\cite{mistral7b}} & 0 & 0 & 0 \\ \hline
\textbf{DeepSeek-R1-7B~\cite{deepseek-coder}}          & 0    & 0      & 0     \\ \hline
\textbf{LLama3.2-3B-Instruct~\cite{llama3p2_3b_instruct}} & 0    & 0      & 0     \\ \hline
\textbf{LLama3.1-8B~\cite{llama3p1_8b}}          & 0    & 0      & 0     \\ \hline
\textbf{GPT-3.5 Turbo~\cite{GPT3.5-turbo}}        & 0.15 & 0.60 & 0.89 \\ \hline
\textbf{GPT4~~\cite{GPT4}}                 & 0.3  & 0.87  & 0.99  \\ \hline
\textbf{GPT-4o-mini~\cite{GPT4o-mini}} & 0.05    &   0.25    & 0.5     \\ \hline
\textbf{Veritas [Ours]}      &  \textbf{1} &  \textbf{1}  & \textbf{1}  \\ \hline
\end{tabular}
}
\end{table}
% 
% 
%\begin{minipage}{\linewidth}

%\end{minipage}
% 
 
\subsection{Case Study 2}

\textbf{Scope.}
In this case study, we will test \emph{the CNF generation capability of Veritas} for circuits.

\textbf{Setting.}
We use our test dataset comprising 132 data points. We ensure that the prompts used in the test dataset to generate the CNFs are not a part of either training or validation data.

\textbf{Results.}
Our fine-tuned model was able to create CNFs for 2 to 4-bit adders, 2 to 4-bit subtractors, $2\times1$ to $16\times1$ multiplexers, and $2\times4$ decoders to  $5\times32$ decoders. We tested this CNF for equivalence against our golden CNFs. We achieved 100\% equivalence with the golden CNFs for adder, subtractor, and multiplexer. We provided a maximum token limit of 1200 for a generation.
 
\subsection{Case Study 3} 
 
\textbf{Scope.}
In our final case study, we try to generate a \emph{fully functional ALU Verilog code} using the CNFs generated by Veritas.

\textbf{Setting.}
Our designed ALU takes two 4-bit inputs, $A$ and $B$, and outputs a 4-bit output $Y$. It comprises a 4-bit adder, a 4-bit subtractor unit, a logical `AND', and a logical `OR'. The output bits of each module are sent to four $4\times1$ MUXes, whose outputs are the final output. The $s_0$ and $s_m$ select lines are used to select the output of the respective operating unit. 

\textbf{Results.}
We use Veritas to generate the CNFs for each module and obtain the corresponding Verilog code for each module. As a post-processing phase, we combine the Verilog codes into a top file according to the design (Figure~\ref{fig:toy_alu}). The ALU generated is syntactically and functionally correct when implemented. Next, we give a detailed comparison with the SoTA. 

\subsection{Comparison with the State-of-the-art} 
We compare the performance of Veritas in \textit{Verilog code generation} with two open-source works, Verigen~\cite{thakur2024verigen} and RTLCoder~\cite{rtlcoder}. We show how Verigen-2B and RTLCoder fail to generate fundamental circuits like adders and subtractors, and eventually, non-trivial circuits like ALUs are built upon these primary modules. 
\subsubsection{\bf Veritas vs Verigen-2B}

We generate four different designs using Verigen-2B. These designs include 4-bit adder, 4-bit subtractor, $16\times1$ multiplexer, and $5\times32$ decoder. We have used temperature = 0.7 and top\_p = 0.9   for these generations. We generate 10 instances of each of these designs using the type of prompt represented in Verigen~\cite{thakur2024verigen}. For example, for 4-bit adder and subtractor, we prompt ``module 4-bit adder''.

\ul{Results on Verigen.}
For the adder design, we observe completions that inherently uses 1-bit adder's instances, which are not defined in the generated Verilog code. Similar observations are seen for the code generated with the subtractors. Hence, these code are not ready to be used in the form they have been generated. 

Although the decoder’s generated code compiled without errors, it failed to implement the intended logic (see Listing~\ref{lst:4bit_sub_Verigen}). Further, it can be seen that for the inputs of 6'b000001 to 6'b001111 (Listing~\ref{lst:4bit_sub_Verigen}), the incorrect bit positions (corresponding to hex) are set in the output. However, for the remaining inputs none of the bits are set. We also tried descriptive prompting, similar to what we use in Veritas on Verigen, but Verigen-2B could not produce a single design, not even partially.

\ul{Results on Veritas.}
We prompted Veritas with 10 prompts comprising the above design and got the correct CNF for all 10 designs. 
We converted each of these CNF formulae into Verilog, validated their functionality, and confirmed 100\% equivalence to the reference designs using the ABC tool~\cite{abc}. 

\subsubsection{\bf Veritas vs RTLCoder-v1.1-gguf-4bit} 
RTLCoder~\cite{rtlcoder} that has been fine‑tuned on a comprehensive dataset of 27,000 Verilog designs, is the top-performing open‑source LLM at the time of writing this paper; however, our work focuses on lighterweight models, implementable with CPUs.
\begin{lstlisting}[caption={Incorrect logic of 6x64 decoder (Combinational) generated by Verigen-2B. The output is incorrect for input 6'b000001 and onward (highlighted in red).}, label={lst:4bit_sub_Verigen}]
§Prompt:§
//module 6x64 decoder (input [5:0] sel, output reg [63:0] out);
§Completion:§
        always @(sel)
        begin
                case(sel)
                        6'b000000: out = 64'h0000000000000001;
                        §6'b000001: out = 64'h0000000000000010;
                        6'b000010: out = 64'h0000000000000100;
                        6'b000011: out = 64'h0000000000001000;
                        ................
                        6'b001111: out = 64'h1000000000000000;
                        6'b010000: out = 64'h0000000000000000;
                        ................
                        6'b111011: out = 64'h0000000000000000;
                        6'b111100: out = 64'h0000000000000000;
                        6'b111101: out = 64'h0000000000000000;
                        6'b111110: out = 64'h0000000000000000;
                        6'b111111: out = 64'h0000000000000000;§
                endcase
        end
endmodule
 \end{lstlisting}
Hence, RTLCoder's lightweight version (size $\approx$ 4 GB), `RTLCoder-v1.1-gguf-4bit', was our next choice.

Following Case Study 3 and the experiments we did with Verigen-2B, we asked RTLCoder to generate a simple 4-bit adder. We specified the input and output and model implementation details adhering the to format provided in their GitHub repository~\cite{RTLCoderGithub}.

\ul{Results for RTLCoder.}
The completion was syntactically correct but functionally not. For the scenario where we added two zeroes as input, the generated Verilog code returned 1 in the carry (see Listing~\ref{lst:4bit_sub_RTLCoder}). Similarly, the 4-bit subtractor's code was functionally incorrect despite being syntactically correct. Next, we tried to generate a $16\times1$ multiplexer, which RTLCoder generated correctly. 
% It is evident from the results that VeriCNF has an edge over the other two thanks to its CNF-based Verilog generation step. 

\ul{Results for Veritas.}
The results clearly demonstrate that Veritas outperforms other two methods due to its CNF‑based Verilog generation process. A pass@1 and pass@5 comparison between Verigen, RTLCoder, and Veritas tested on similar prompt has been provided in Table~\ref{tab:comparison}. 
\subsection{Summary} 
Through Case Studies 1, 2, and 3, we evaluated the potency of employing Veritas to synthesize CNFs and reconstruct deterministic Verilog representations.

Case Study 1 analyzed the proficiency of various open-source and proprietary models in CNF generation for basic logic gates. The analysis revealed that most models faltered in generating correct CNFs, even for rudimentary constructs. Notably, GPT-4 consistently derived valid CNFs for buffers and produced a correct XOR gate CNF only once, positioning it closest to Veritas regarding \emph{pass@k} performance.
Case Study 2 illustrated Veritas’ capacity to infer CNFs for foundational RTL designs, including adders, subtractors, multiplexers, and decoders. These CNFs were seamlessly converted into Verilog code that proved operational and devoid of syntactical and logical errors in Case Study 3. Those modules were further synthesized and integrated to fabricate a working ALU.

Finally, in our comparative evaluation against contemporary SoTA, we examined Verigen-2B and RTLCoder, two open-sourced RTL generative frameworks. However, both failed to produce functional constructs for critical components such as adders, subtractors, and decoders, compromising the viability of using their outputs to assemble a simple ALU.

\begin{table}[tb]
\centering
%\resizebox{0.9\columnwidth}{!}{%
%\footnotesize
\begin{tabular}{|l|c|c|c|}
\hline
\textbf{Models} & \textit{\textbf{Verigen}} & \textit{\textbf{RTLCoder}} & \textbf{Veritas} \\ \hline
\textbf{pass@1} & 0.2  & 0.35 & 1 \\ \hline
\textbf{pass@5} & 0.77 & 0.91 & 1 \\ \hline
\end{tabular}%
%}
\caption{A comparison with the SoTA open-source lightweight LLM-based Verilog code generators.}
\label{tab:comparison}
\end{table}

\section{Conclusions}
 
This paper introduced Veritas, a novel CNF-guided synthesis methodology that leverages lightweight LLMs to generate correct-by-construction Verilog. Unlike traditional HLS approaches, which often struggle to scale to real-world designs, or conventional LLM-driven methods that require iterative debugging, Veritas directly outputs formal CNF specifications representing circuit functionality. These specifications are deterministically converted to Verilog, ensuring functional correctness without external verification.
Post peer-review, we will release our method in full.

By fine-tuning the lightweight, open-source CPU-deployable LLama-3.2-3B-Instruct model on standard RTL components, we demonstrated significant reliability and accuracy in synthesizing hardware designs, including non-trivial components such as ALUs.

Comparative studies against state-of-the-art lightweight models like Verigen-2B and RTLCoder-7B highlight Veritas' enhanced capability to produce correct RTL designs on the first attempt.

Finally, our CNF-based pipeline merges hardware specification and implementation, significantly streamlining the hardware development process by eliminating costly and time-consuming verification stages inherent in existing LLM-assisted workflows. 

A potential future direction is to explore fine-tuning larger LLMs on more extensive benchmarks to enhance generation capabilities and support increasingly complex designs.
% 
%\section{Acknowledgments}

% \begin{acks}
% Acknowledgement goes here.
% \end{acks}

%\newpage
\bibliographystyle{ACM-Reference-Format}
%\bibliography{Ref}

% \appendix

\end{document}